\documentclass{desyproc}

\usepackage{subfigure}
\usepackage[sort&compress,square]{natbib}

\begin{document}
\title{\vspace{-3cm}
\hfill{\small{IPPP/08/68; DCPT/08/136}}\\[2cm]
The Physics Case for Axions, WIMPs, WISPs and Other Weird Stuff}

\author{{\slshape Joerg Jaeckel}\\[1ex]
Institute for Particle Physics and Phenomenology, Durham University, Durham DH1 3LE, UK\\
}

\contribID{jaeckel\_joerg}

\desyproc{DESY-PROC-2008-02}
\acronym{Patras 2008} 

\maketitle

\begin{abstract}
We argue that there exists an excellent `physics case' motivating
the search for axions, WIMPs, WISPs and other phenomena testable at
low energies. This physics case arises from both experimental and observational evidence
as well as the desire to test theoretical model building.
\end{abstract}

\section{Introduction -- Hints for new physics}
Over the years both theoretical as well as experimental evidence has accumulated that strongly suggests the existence of physics beyond the current standard
model of particle physics (SM). The Large Hadron Collider currently starting up at CERN will test many of the ideas for such physics beyond
the standard model (BSM) and hopefully
will provide us with a wealth of new information.
In this note we argue that there is also a very good motivation to search for new physics in low energy experiments that can provide us with powerful
complementary information on currently open questions and in particular on how the standard model is embedded into a more fundamental theory.

Let us begin by briefly repeating some of the main reasons why we believe that there must be physics beyond the standard model.

On the theoretical side there are a number of deficiencies in the SM. Some of them could be just aesthetic defects but some may go deeper.
First of all the SM has a relatively large number ${\mathcal{O}}(30)$ free parameters that cannot be determined from theory alone but must be measured experimentally.
Although this does not indicate an inconsistency of the theory it certainly is not in line with the hope that a fundamental theory of everything should have
very few, possibly only $1$ or even $0$, free parameters.
Moreover, some of the parameters seemingly need to require an enormous degree of finetuning or appear unnaturally small. Well known examples are the Higgs mass
but also the $\theta$ parameter of QCD (which must be extremely small in order not to be in conflict with the observed smallness of strong CP violation).
Another dissatisfying feature is that gravity is not incorporated into the SM but rather treated as a separate part. This is not just an aesthetic defect but also an expression
of the fact that the quantization of gravity is still not (fully) understood.
Finally, strictly speaking the SM will most likely not be valid up to arbitrary high energy scales. On the one hand this is due to our current inability
to properly quantize gravity. But even the non-gravity parts are probably encountering problems in the form of Landau poles (places where the coupling becomes infinite)
in the QED sector (at a very high scale much beyond the Planck scale) but probably also in the Higgs sector (where the problem is much more immediate and will occur
at scales much below the Planck scale - depending on the Higss mass possibly even not much above the electroweak scale).

Next there are quite a few phenomena which are experimentally well established but for which there is no good explanation within the standard model.
The most shocking of which is probably the realization that most of the matter and energy in the universe actually is not made up of SM particles.
Cosmological and astrophysical observations give strong evidence that about 70 \% of the energy in the universe is dark
energy and another 25 \% is dark matter~\cite{Komatsu:2008hk}.
These are things that simply do not appear in the current SM (although they could be accommodated see, e.g., \cite{Davoudiasl:2004be}).
But even within the standard model there are things which are experimentally well established but for which a good explanation is lacking. These are, e.g., the
existence of three generations of SM particles, the mass
hierarchies for the SM particles and the small parameters such as, e.g., the already mentioned $\theta$ parameter~\cite{Ramsay}.
The latter is, of course, a repetition of some
of the problems already mentioned as `theoretical' problems showing that they actually arise from experimental results.

Finally, there is the direct experimental evidence for BSM physics. At the moment most of this is still relatively circumstantial but it definitely demonstrates
that low energy experiments can provide information on BSM physics as well as opening new directions which can be explored (or close others).
Examples are the deviation~\cite{Bennett:2004pv} of the muon $(g-2)$ from the SM expectation, the excess in the event rate of the DAMA~\cite{Bernabei:2008yi} experiment and
the PVLAS anomaly~\cite{Zavattini:2005tm} (which has
been retracted~\cite{Zavattini:2007ee} but, as we will see, has inspired a lot of fruitful experimental and theoretical activity).



\section{Bottom-up/phenomenological arguments}
In this section we will present several examples for physics at the low energy frontier that arise from more phenomenological arguments - a line of thought
that could be called `bottom-up' and that follows a hands-on approach on fixing problems step by step.

Axions are a good example for this approach~\cite{Peccei:1977hh}. The extreme smallness of the $\theta$-angle is unexplained in the standard model. This can be solved by introducing a new
symmetry - the Peccei Quinn symmetry.  As a consequence one predicts a pseudo-Goldstone boson, the axion. This is already a good motivation to experimentally
search for the axion, for example in light shining through a wall experiments~\cite{regeneration,Cameron:1993mr,Ehret:2007cm}, laser polarization
experiments (as, e.g., PVLAS)~\cite{Zavattini:2005tm,Chen:2006cd,Zavattini:2007ee} or axion helioscopes~\cite{Zioutas:2004hi}.
The case for this search is then strengthened by the finding that the axion is also a valid candidate for dark matter~\cite{Preskill:1982cy}.
This prediction, however, not only strengthens
the physics case for searching axions but it also opens new ways to do so. One can search for axion dark matter, for example using resonant
cavity techniques~\cite{Sikivie:1983ip}
or looking in the sky for axions decaying into photons.

Another example are WIMPs (for a review see~\cite{Jungman:1995df}). As a solution for the hierarchy problem in the SM one can, again, introduce a symmetry: SUSY.
Introducing SUSY leads to many new particles, notably the heavy supersymmetric partners of the SM particles,
which are weakly interacting and massive, i.e. WIMPs. Some of them are good candidates for dark matter. Again good motivation to perform a WIMP search.
Another incentive is that SUSY also allows to explain the deviation of the muon $(g-2)$ from its SM value that was already mentioned in the introduction.
SUSY might be discovered at a collider such as LHC. Such an experiment may even find a dark matter candidate.
But in order to know that such a candidate really makes up all or most of
the dark matter, i.e. if it was produced in sufficient quantities, one needs the low energy WIMP searches~\cite{Bernabei:2008yi,Probst:2002qb} which therefore give us crucial
information.

The PVLAS anomaly which was in contradiction to the SM expectation led to the introduction of several types of
WISPs (weakly interacting slight (or sub-eV) particles). To check their
result and to search for these WISPs the PVLAS group then improved their apparatus finding that the original result was
probably an artifact of the apparatus~\cite{Zavattini:2007ee}.
However, this is not the end of the story. The introduction of WISPs also led people to realize that there is a large amount of unexplored
parameter space for new physics that (e.g., due to the extremely weak interactions involved) cannot be tested
in conventional colliders~\cite{Masso:2005ym}~\footnote{Astrophysical arguments are, however, a different matter. For an overview over pre-PVLAS work in this
direction see, e.g.,~\cite{Raffelt:1996}.}.
Yet new ideas how to access this parameters space in low energy experiments and observations
have been put forward~\cite{Dupays:2005xs}.
Moreover, it was (re-)discovered that the extremely weak interactions of WISPs are often connected to very high energy scales $\gtrsim 10^{5}$~GeV, in some
cases even as high as the string or the even Planck scale $\sim 10^{18}$~GeV. Showing that the new
and improved low energy experiments can give us complementary information on very high energy physics.

\section{Top-down/theory arguments}
Instead of taking small steps and fixing the problems, in the process often creating a more and more baroque model, one can also go back to the drawing
board and rethink the very principles on which the original model was based. One such attempt (among others) is string theory.
One of the main motivations for string theory is to unify the SM with gravity. To achieve this point particles are replaced by extended strings.
Currently string theory is not yet in a state where it provides a first principle derivation of the SM and corrections to the same.
Nevertheless, it has a variety of general features that suggest avenues for model building and also specific phenomena.

One such general feature is that for consistency string theory likes SUSY. Following the arguments from the previous section SUSY provides a good
physics case for WIMP searches. Accordingly string theory strengthens the physics case for such searches.

Another property of string theory is that in order to be consistent it needs the existence of extra (space) dimensions. In order to be in
agreement with observation all except the well known three have to be compactified. However, compactification leaves
its traces\footnote{If the size of the extra dimension is large enough there could actually be a very direct consequence: the inverse square law of the gravitational
force would be modified. This, too, can be tested in low energy experiments~\cite{Adelberger:2003zx}.}. Shape and size deformations
of the compactified dimensions correspond to scalar fields, so-called moduli. These could be very light (it is actually often difficult to give them any mass at all)
and provide excellent WISP candidates (and may also be searched for in fifth force experiments~\cite{Adelberger:2003zx}). In a similar manner also various types of axions appear in string theory (see, e.g., \cite{Svrcek:2006yi}).
The physics of these particles (e.g. the small size of their interactions) is inherently linked to the string scale.
Hence, suitable low energy experiments searching for such
WISPs may give us the opportunity to probe the fundamental theory and its associated fundamental energy scale.

String theory also tends to have whole sectors of extra matter in addition to the ordinary SM matter. This matter often lives in so-called hidden sectors which have
only extremely weak interactions with the SM particles. Accordingly particles in these sectors may avoid detection in collider experiments even if they are
light, i.e. these hidden sector provide good candidates for WISPs\footnote{If the hidden sector particles are somewhat heavier they can
also be WIMP candidates~\cite{Kors:2004dx} (which in some cases can also be searched for at colliders~\cite{Kors:2004dx,Coriano:2008wf}).
However, this also depends on how strict one takes the `Weak' in the name WIMP. Often it is constraint
to be the weak of electroweak. Then hidden sector particles cannot really be WIMPs.}. Typical
WISP candidates arising from such hidden sectors are extra `hidden' U(1) gauge bosons and `hidden' matter charged under
those U(1)s~\cite{Batell:2005wa,Dienes:1996zr}\footnote{The hidden U(1)'s can interact with the SM via a kinetic or mass mixing with the
ordinary photon~\cite{Okun:1982xi,Holdom:1985ag}.
This kinetic
mixing can then also lead to a small electric charge for the hidden matter~\cite{Holdom:1985ag}.}.
In many models these hidden sectors are located at a different place in the extra dimension than the SM sector\footnote{An alternative to truly hidden
sectors are sectors with hyperweak interactions~\cite{Burgess:2008ri}.
Although in these models the new particles can have tree-level interactions with the standard model particles these are extremely weak. Effectively they
are diluted because the hyperweak sector extends more
into the extra dimensions. Accordingly these, too, contain more information about the global structure.}. Accordingly
searching and testing these hidden sectors can give us crucial global information on the compactification that can hardly be obtained from collider experiments
which probe the local structure that has relatively strong interactions.

Finally, string theory also motivates some surprising things. In particular, some models predict non-commutativity and other Lorentz symmetry
violating effects~\cite{Kostelecky:1988zi}. This then can also be tested in low energy experiments and observations such as, e.g., comparing the spectrum of hydrogen and
anti hydrogen atoms~\cite{Lehnert:2006gw} or by observing if light from gamma-ray bursts arrives at (slightly) different times
depending on its polarization~\cite{Kostelecky:2006ta}.
These experiments and observations (see \cite{Kostelecky:2002hh} for an overview) again provide an ultra high precision that can then
give us insights into the fundamental theory at very high energy scales.

\section{Conclusions}
Both the phenomenological bottom-up and the more theory oriented top-down approach provide an excellent physics case motivating further experiments at the low-energy
frontier such as searches for axions, WIMPs and WISPs and other interesting effects such as Lorentz violation.
These phenomena are often connected to energy scales much higher than those reachable in near future accelerators.
They provide experimental access to hidden sectors that may contain crucial information on the underlying global structure of a more
fundamental theory. Moreover, they give us reasons to challenge and experimentally check basic assumptions as, e.g., Lorentz symmetry.
In conclusion, low energy but high precision experiments provide crucial complementary information to uncover the nature of a more fundamental theory beyond the standard model.

\noindent
{\em Final Note:} Many of the things mentioned in these proceedings have been intensely discussed at the
 \emph{Brainstorming and Calculationshop: The Physics Case for a Low Energy Frontier of Fundamental Physics} held at DESY in June 2008.
 More details from this collective effort can be found soon on
http://alps-wiki.desy.de/e13/e42.
\section*{Acknowledgements}
The author wishes to thank the organizers of the \emph{4th Patras Workshop on Axions, WIMPs and WISPs} for a very enjoyable and stimulating meeting.
Moreover, he is indebted to all participants of the \emph{Brainstorming and Calculationshop: The Physics Case for a Low Energy Frontier of Fundamental Physics} for
a great meeting on the subjects discussed in these proceedings.

\begin{footnotesize}

\end{footnotesize}


\end{document}